\documentclass[reprint,superscriptaddress,amsmath,amssymb, 11pt,onecolumn,aps,prl,showpacs]{revtex4-1}

\usepackage[english]{babel}
\usepackage{multirow}
\usepackage[thinspace]{SIunits}
\usepackage{xspace}
\usepackage{graphicx}
\graphicspath{{figures//}} 
\usepackage{dcolumn}
\usepackage{bm}
\usepackage{color}
\usepackage{wrapfig}
\usepackage{hyperref}
\usepackage{array}
\usepackage{booktabs}
\usepackage{multirow}
\hypersetup{
    unicode=false,          
    pdftoolbar=true,        
    pdfmenubar=true,        
    pdffitwindow=false,     
    pdfstartview={FitH},    
    pdftitle={My title},    
    pdfauthor={Author},     
    pdfsubject={Subject},   
    pdfcreator={Creator},   
    pdfproducer={Producer}, 
    pdfkeywords={keyword1} {key2} {key3}, 
    pdfnewwindow=true,      
    colorlinks=true,       
    linkcolor=blue,          
    citecolor=blue,        
    filecolor=blue,      
    urlcolor=blue,       
    plainpages=false        
}

\newcommand{\etal}{\textit{et al.}\xspace}

\newcommand{\Tcdw}{$T_{\mathrm{{CDW}}}$\xspace}

\newcommand{\CVS}{\mbox{CsV$_3$Sb$_5$}\,\xspace}
\newcommand{\Alg}{$A_{\rm {1g}}$\,\xspace}

\newcommand{\Elg}{${E_{1g}}$\,\xspace}
\newcommand{\Ezg}{$E_{2g}$\,\xspace}

\newcommand{\wn}{\rm{cm}$^{-1}$\,}

\begin{document}

\title{\Large Anharmonic Strong-Coupling Effects at the Origin of the Charge Density Wave in CsV$_3$Sb$_5$}

\author{G. He}
\altaffiliation{These authors contributed equally to the work.}
\affiliation{Walther Meissner Institut, Bayerische Akademie der Wissenschaften, Garching 85748, Germany}

\author{L. Peis}
\altaffiliation{These authors contributed equally to the work.}
\affiliation{Walther Meissner Institut, Bayerische Akademie der Wissenschaften, Garching 85748, Germany}

\author{E. Cuddy}
\altaffiliation{These authors contributed equally to the work.}
\affiliation{Department of Materials Science and Engineering, Stanford University, Stanford, California 94305, USA}
\affiliation{Stanford Institute for Materials and Energy Sciences, SLAC National Accelerator Laboratory and Stanford University,
2575 Sand Hill Road, Menlo Park, California 94025, USA}

\author{Z. Zhao}
\affiliation{Beijing National Laboratory for Condensed Matter Physics, Institute of Physics, Chinese Academy of Sciences, Beijing 100190, China}

\author{D. Li}
\affiliation{Beijing National Laboratory for Condensed Matter Physics, Institute of Physics, Chinese Academy of Sciences, Beijing 100190, China}

\author{R. Stumberger}
\affiliation{Walther Meissner Institut, Bayerische Akademie der Wissenschaften, Garching 85748, Germany}
\affiliation{Fakult\"at f\"ur Physik E23, Technische Universit\"at M\"unchen, Garching 85748, Germany}

\author{B. Moritz}
\affiliation{Stanford Institute for Materials and Energy Sciences, SLAC National Accelerator Laboratory and Stanford University,
2575 Sand Hill Road, Menlo Park, California 94025, USA}

\author{H. T. Yang}\email[Corresponding author: ]{htyang@iphy.ac.cn}
\affiliation{Beijing National Laboratory for Condensed Matter Physics, Institute of Physics, Chinese Academy of Sciences, Beijing 100190, China}
\affiliation{School of Physical Sciences, University of Chinese Academy of Sciences, Beijing 100049, China}

\author{H. J. Gao}
\affiliation{Beijing National Laboratory for Condensed Matter Physics, Institute of Physics, Chinese Academy of Sciences, Beijing 100190, China}
\affiliation{School of Physical Sciences, University of Chinese Academy of Sciences, Beijing 100049, China}

\author{T. P. Devereaux}\email[Corresponding author: ]{tpd@slac.stanford.edu}
\affiliation{Department of Materials Science and Engineering, Stanford University, Stanford, California 94305, USA}
\affiliation{Stanford Institute for Materials and Energy Sciences, SLAC National Accelerator Laboratory and Stanford University,
2575 Sand Hill Road, Menlo Park, California 94025, USA}
\affiliation{Geballe Laboratory for Advanced Materials, Stanford University, Stanford, California 94305, USA}

\author{R. Hackl}\email[Corresponding author: ]{Hackl@tum.de}
\affiliation{Walther Meissner Institut, Bayerische Akademie der Wissenschaften, Garching 85748, Germany}
\affiliation{Fakult\"at f\"ur Physik, Technische Universit M\"unchen, Garching 85748, Germany}
\affiliation{IFW Dresden, Helmholtzstrasse 20, Dresden 01062, Germany}

\date{\today}

\begin{abstract}

The formation of charge density waves (CDW) is a long-standing open problem particularly in dimensions higher than one \cite{Peierls:1955, Varma:1983, Eiter:2013}. Various observations in the vanadium antimonides discovered recently \cite{Ortiz:2019,Ortiz:2020}, such as the missing Kohn anomaly in the acoustic phonons \cite{Li:2021,Xie:2022} or the latent heat at the transition $T_{\rm CDW} = 95$\,K \cite{Ortiz:2019, Ni:2021}, further underpin this notion. Here, we study the Kagome metal CsV$_3$Sb$_5$ using polarized inelastic light scattering. The electronic energy gap $2\Delta$ as derived from the redistribution of the continuum is much larger than expected from mean-field theory and reaches values above 20 for $2\Delta/k_{\rm B}T_{\rm CDW}$. The \Alg phonon has a discontinuity at $T_{\rm CDW}$ and a precursor starting 20\,K above $T_{\rm CDW}$. Density functional theory qualitatively reproduces  the redistribution of the electronic continuum at the CDW transition and the phonon energies of the pristine and distorted structures. The linewidths of all \Alg and \Ezg phonon lines including those emerging below \Tcdw were analyzed in terms of anharmonic symmetric decay \cite{Klemens:1966} revealing strong phonon-phonon coupling. In addition, we observe two CDW amplitude modes (AMs): one in \Alg symmetry and one in \Ezg symmetry.  The temperature dependence of both modes deviates from the prediction of mean-field theory. The \Alg AM displays an asymmetric Fano-type lineshape, suggestive of strong electron-phonon coupling. The asymmetric \Alg AM, along with the discontinuity of the \Alg phonon, the large phonon-phonon coupling parameters and the large gap ratio, indicate the importance of anharmonic strong phonon-phonon and electron-phonon coupling \cite{Varma:1983} for the CDW formation in \CVS.

\end{abstract}
\maketitle

\section{Introduction}
Lattices of magnetic ions having regular triangular coordination are characterized by multiple ordering phenomena including ferromagnetism, frustrated antiferromagnetism, density waves  and superconductivity (SC).  These lattices attracted a lot of attention not only for the magnetism but also for the specific band structure being characterized by a Dirac dispersion and Weyl nodes induced by spin-orbit coupling. As a typical example, the vanadium-antimony compound class $A$V$_3$Sb$_5$ ($A=$\,K, Rb, Cs) forming a Kagome lattice with hexagons and triangles alternating was discovered recently \cite{Ortiz:2019,Ortiz:2020, Ortiz:2021,Yin:2021}.
The V-Sb Kagome layers are separated by Sb honeycomb-like layers and alkali monolayers as shown in Fig. \ref{fig:lattice}~\textbf{a}. At low temperature, charge density waves (CDW) and SC may occur. The focus here is placed on the CDW transition forming a $2\times2\times2$ superlattice at \Tcdw in the 100-kelvin range which may be driven by an unconventional mechanism beyond electron-phonon interaction. Rather, the proximity to a Van Hove singularity close to the Fermi surface is considered responsible for the instability \cite{Li:2021}.

Obviously, the ordering vector \textbf{Q} connects two M points (see the green arrow in Fig. \ref{fig:lattice}~\textbf{d}) in agreement with the electronic structure predicted theoretically \cite{Zhao:2021} and observed by angle-resolved photoemission spectroscopy (ARPES) \cite{Kang:2022} and scanning tunneling spectroscopy (STS) \cite{Zhaoh:2021}. Yet, the meaning of the observed energy scales is still controversial. ARPES \cite{Wang:2021} and STS \cite{Jiang:2021, Liang:2021} find a gap at 20\,meV and thus just one-fourth of the scale observed by infrared spectroscopy \cite{Zhou:2021, Uykur:2021}. More recent ARPES measurements disclose that the small gap may originate from massive Dirac points \cite{Nakayama:2021}, and a larger CDW gap may open at the $M$ points \cite{Nakayama:2021, Luo:2022} and corresponds to a ratio $2\Delta/k_B T_{\mathrm{{CDW}}} \sim 20$ far beyond the weak coupling prediction of 3.53. Complementary to spectroscopic methods thermodynamic studies indicate a divergence in the heat capacity being more compatible with a first-order rather than a second-order transition as usually expected for a CDW \cite{Ortiz:2020}.

There are various experimental methods that can be used to attack this issue. One may look for anomalies close to the ordering vector \textbf{Q} in the acoustic phonon branches using either neutron \cite{Xie:2022} or inelastic x-ray scattering \cite{Li:2021}. This search has been unsuccessful so far, and the conclusion reached is that either k-dependent electron-phonon coupling or electron-electron interaction is the origin of CDW ordering. Alternatively, optical phonons displaying renormalization effects at \Tcdw \cite{Wulferding:2022,Wu:2022,LiuG:2022} or Fano-type line shapes may indicate strong electron-phonon coupling. In addition to phonons, oscillations of both the amplitude and the phase of the order parameter are expected for a CDW system \cite{Lee:1993}. For symmetry reasons, Raman scattering and time resolved techniques project the amplitude mode (AM) directly thus tracking the CDW phase transition \cite{Grasset:2019, Chen:2017, Lavagnini:2010, Wu:2022,LiuG:2022}. In weak-coupling systems, the AMs are expected to have a symmetric Lorentzian line-shape \cite{Lavagnini:2010, Grasset:2019} with increasing width upon approaching \Tcdw from below. It is not clear
which effect on the AM may be expected if the coupling increases substantially. Finally, the CDW electronic gap is accessible by light scattering.

In this paper, we address the open questions as to the states involving the formation of CDW order, including the size and momentum dependence of the electronic gap, the renormalization of phonons, and the evolution of collective modes, by investigating the temperature and polarization dependent inelastic light scattering response in \CVS. In particular, in contrast to the AMs found in most of other well-known CDW materials, we found the \Alg AM to be 
asymmetric in \CVS, exhibiting a strong Fano resonance between the conduction electrons and the AM. These results along with the strong anharmonic decay of the two prominent Raman-active phonons and most of the CDW-induced phonons highlight the importance of a cooperation between strong phonon-phonon and electron-phonon coupling in the formation of CDW in \CVS.


\section{Results}

\begin{figure}[ht]
   \includegraphics[width=0.5\textwidth]{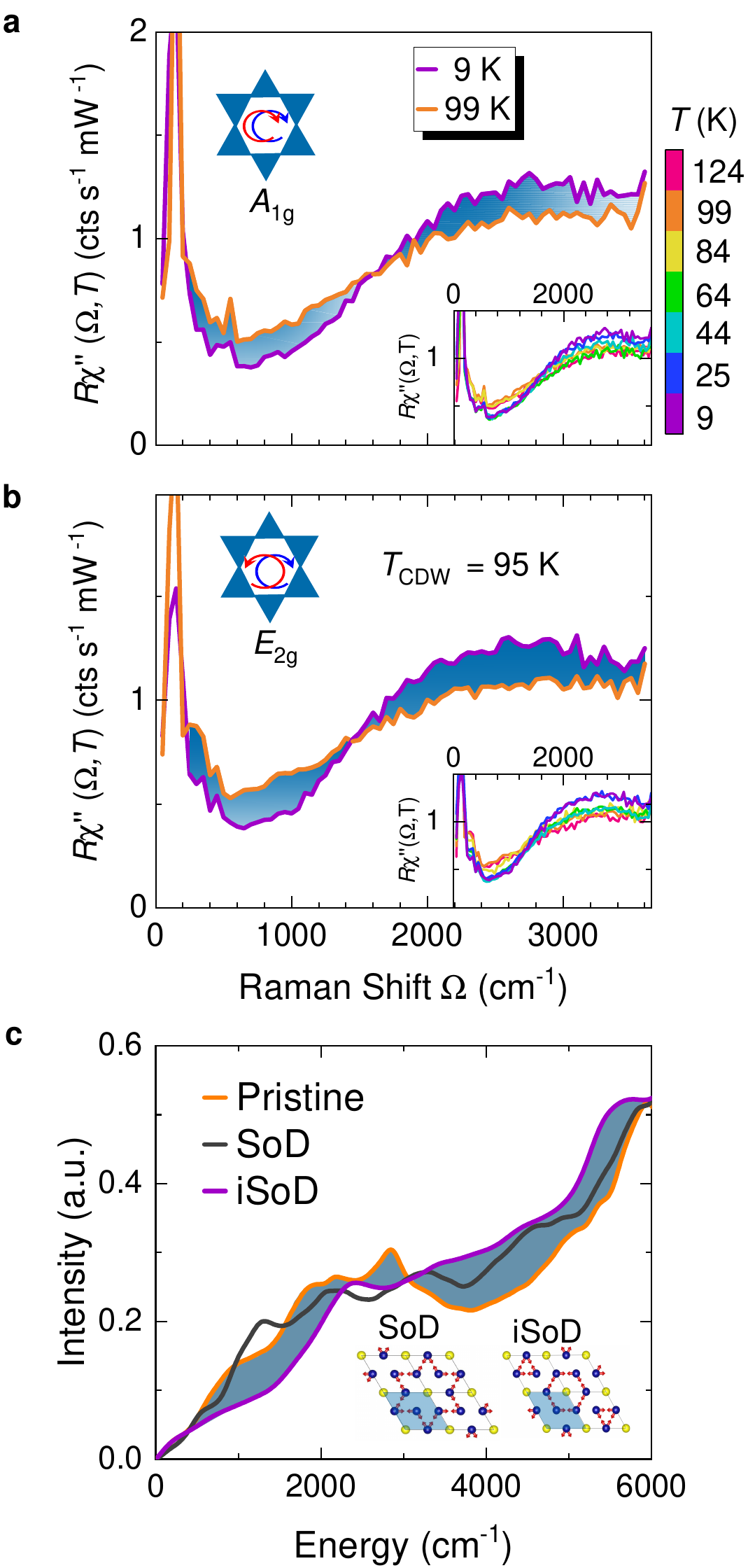}
   \caption{\textbf{CDW gap excitations in both \Alg and \Ezg symmetry.} \textbf{a}
 and \textbf{b} Raman response above and below \Tcdw in \Alg and \Ezg symmetry. The redistribution of the spectral weight is highlighted by the cyan areas. The insets show the detailed temperature-dependent evolution of the spectra, with temperature indicated by the right colorbar. \textbf{c} The FS-integrated electronic response is calculated from DFT using the pristine (black), Star of David (SoD, blue), and inverse Star of David (iSoD, red) lattice. The loss and gain of intensity between the response in the pristine and iSoD-distorted lattice are highlighted by cyan areas. Insets: The SoD and iSoD distortion in the  V-Sb layer. The blue shaded area shows the pristine unit cell. The unit cell below \Tcdw is twice as large.}
   \label{fig:cdwgap}
\end{figure}

Figure \ref{fig:cdwgap} shows the Raman spectra of \CVS above and below \Tcdw for \Alg and \Ezg symmetry in the range from 50 to 3600\,\wn. There is a redistribution of spectral weight from below to above the intersection points at approximately 1\,400 and 1\,540\,\wn in the \Ezg and the \Alg spectra, respectively. The redistribution of the spectral weight is well reproduced for different laser excitations (see Supplementary Materials C for details). The gradual redistribution of spectral weight is a signature of the opening of an energy gap in the electronic excitation spectrum of a CDW system \cite{Eiter:2013}. Upon warming, the amplitude of the redistribution decreases and disappears completely above \Tcdw as shown in the insets of Fig. \ref{fig:cdwgap} \textbf{a} and \textbf{b}. The gap energy is estimated to be close to the intersection points at 87.5 - 96.2\,meV yielding a gap ratio of $2\Delta/k_{\rm B} T_{\rm CDW} \sim 21 - 23$, similar to the results of IR spectroscopy and recent ARPES measurements \cite{Zhou:2021, Uykur:2021, Nakayama:2021, Luo:2022}. As opposed to the IR results, we do neither observe a significant temperature dependence of the intersection points nor the peak energies.

The CDW gap is qualitatively reproduced by the DFT simulations as presented in Fig. \ref{fig:cdwgap} \textbf{c}. The electronic response of \CVS is approximated for pristine-, SoD- and iSoD-distorted-lattices (see the insets of Fig. \ref{fig:cdwgap} \textbf{c}) using the joint density of states (see Methods and Supplementary Materials G for more details). Compared to the response in the pristine-lattice, a reduction in spectral weight is found in the iSoD distorted lattice below 2000 \wn, which can barely be seen in the SoD lattice. This result, together with  the electronic Raman spectra, suggests an iSoD distortion in \CVS to have minimum energy. The mismatch in energy between DFT calculations and the experiments may be reconciled by considering a renormalization factor of approximately 1.5. 

\begin{figure}
   \includegraphics[width=0.6\textwidth]{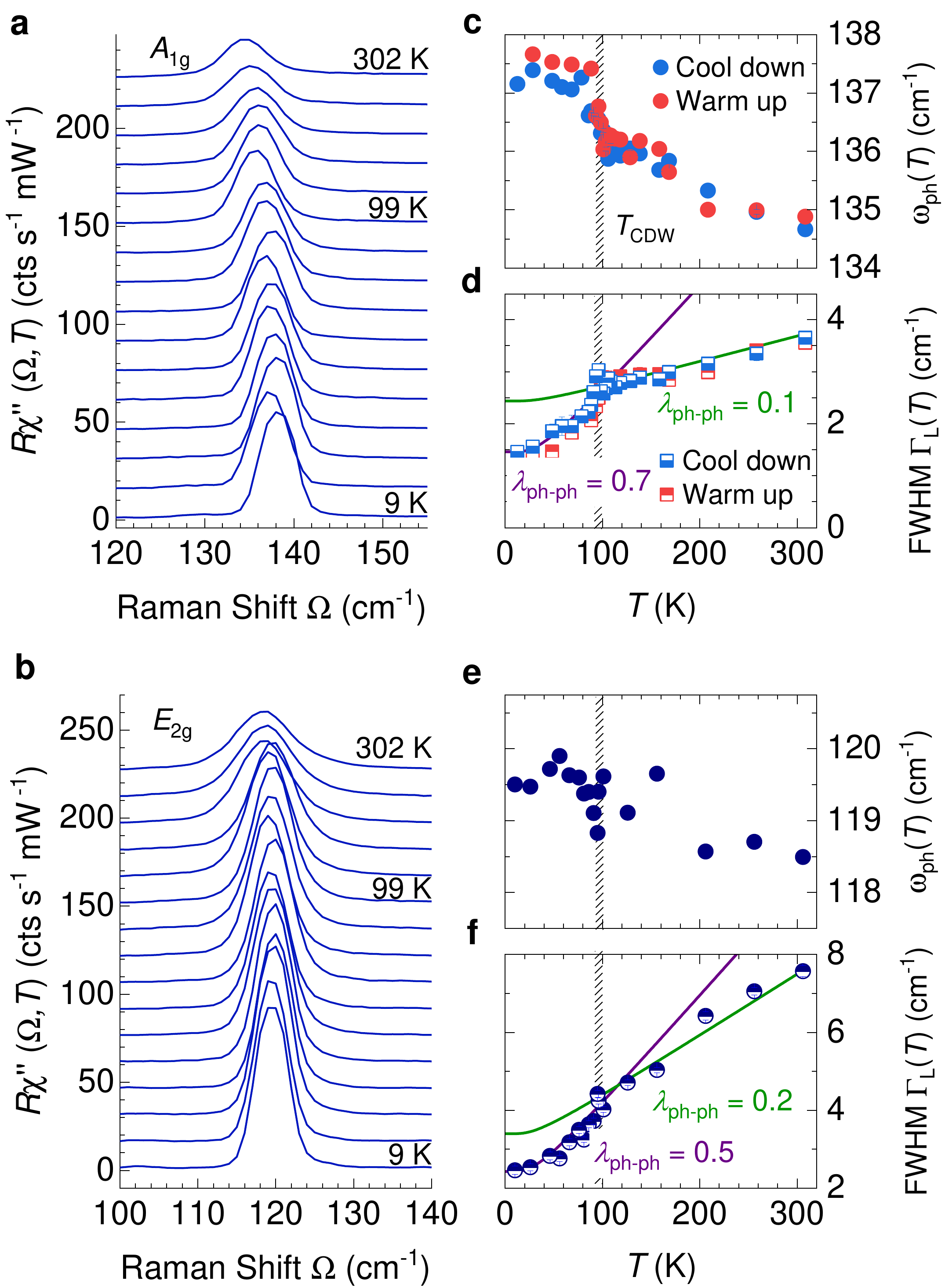}
    \caption{\textbf{Phonon renormalization at \Tcdw.} \textbf{a} and \textbf{b} The temperature evolution of the $A_{1g}$ and the $E_{2g}$ phonon lines was measured at 9, 25, 45, 55, 65, 75, 80, 85, 89, 95, 99, 124, 154, 203, 253, 302\,K, respectively. For clarity, the spectra measured above 9\,K are consecutively offset by 15 cts/(s mW). \textbf{c}-\textbf{f} Phonon energies $\omega_{\rm ph}$ and Full-Width-Half-Maximum (FWHM) of the two lines are derived from a Voigt fit (see Methods). \textbf{c} The $A_{1g}$ mode exhibits a slight hardening right below \Tcdw and a discontinuity at \Tcdw. A hysteresis may exist below \Tcdw. \textbf{e} The temperature dependence of the energy in $E_{2g}$ symmetry is very weak. There may be an anomaly directly at \Tcdw. \textbf{d} and \textbf{f} The temperature dependence of the phonon line widths ($\Gamma_L$) of the $A_{1g}$ and the $E_{2g}$ phonons. The data were fitted separately below and above \Tcdw using an anharmonic model \cite{Klemens:1966} (see Supplementary Materials E for details). There are obvious slope changes at \Tcdw for both the $A_{1g}$ and the $E_{2g}$ phonons.}
   \label{fig:renormalization}
\end{figure}

Two prominent Raman-active phonon lines are observed saturating at 137.5~\wn and 119.5~\wn in the zero-temperature limit for \Alg and \Ezg symmetry, respectively, as shown in Fig. \ref{fig:renormalization}. They have previously been identified by Raman scattering \cite{chen:2021, Wulferding:2022, Wu:2022, LiuG:2022}. 

Both the \Alg and the \Ezg phonons show significant renormalization effects at \Tcdw (see Fig. \ref{fig:renormalization} ). Upon cooling, the \Alg phonon has a precursor starting 20 K above \Tcdw, changes discontinuously at \Tcdw and saturates below (see Fig. \ref{fig:renormalization} \textbf{c}). This discontinuity, a frequency hardening of approximately 0.5\% is replicated in DFT simulations (Details can be found in Table II. of Supplementary Materials G).

Although the energies observed upon cooling and heating do not exactly coincide the hysteresis cannot be considered significant. Thus there is no support for a first-order phase transition from 
the \Alg phonon. The energy of the \Ezg line does not exhibit significant changes across \Tcdw (see Fig. \ref{fig:renormalization} \textbf{e}). A weak dip exactly at \Tcdw may exist but our resolution is not sufficient to have enough confidence. In either symmetry, the line width exhibits a kink at \Tcdw and decreases faster below \Tcdw than above (see Fig. \ref{fig:renormalization} \textbf{d} and \textbf{f}). There may be a weak enhancement of the line width right at the transition.

\begin{figure*}[ht]
   \includegraphics[width=0.9\textwidth]{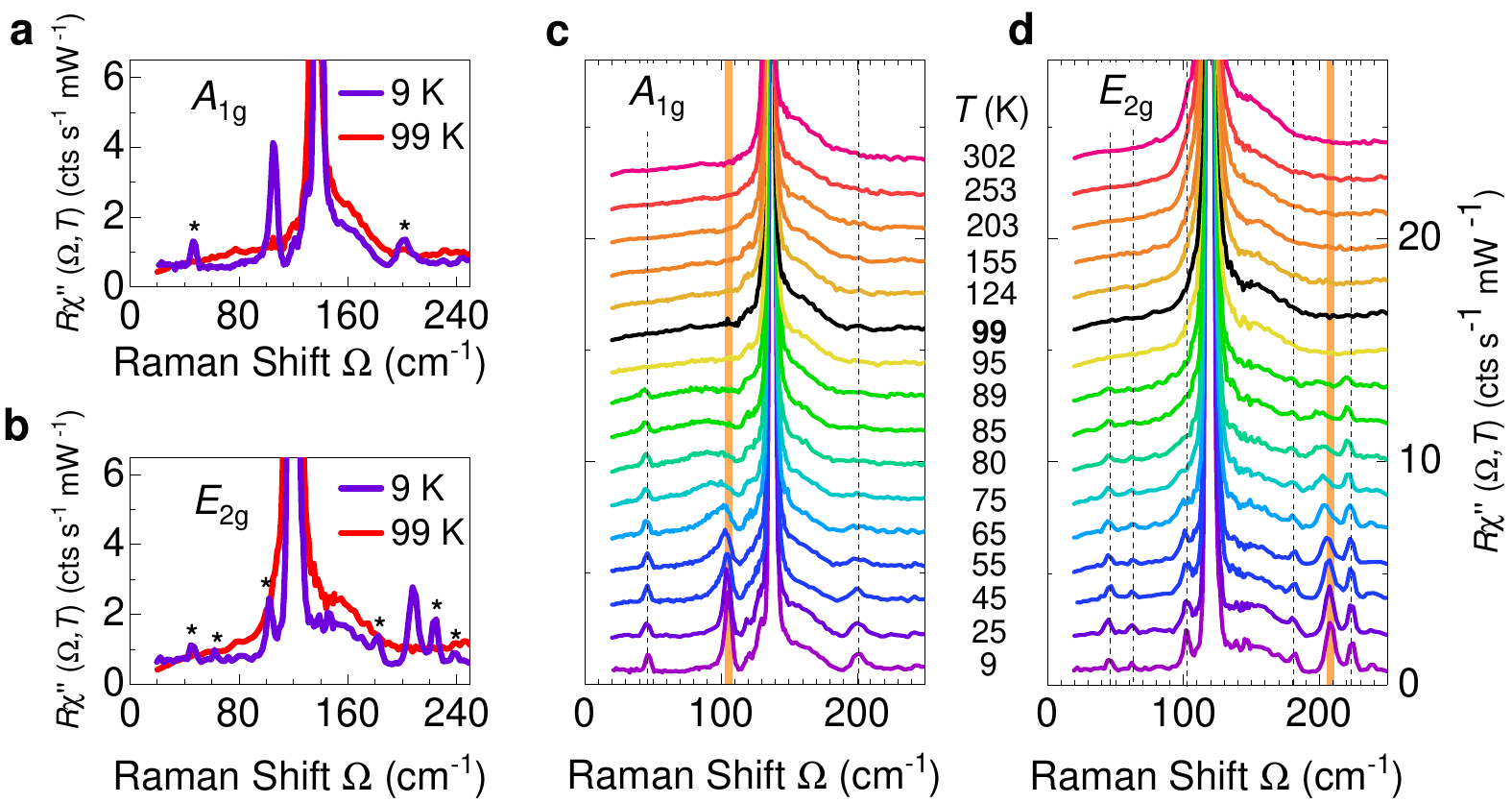}
   \caption{\textbf{Zone folded phonons and amplitude mode in \CVS.} \textbf{a} and \textbf{b} Raman spectra of \CVS below and above \Tcdw in \Alg and \Ezg symmetry. Below \Tcdw, several additional peaks appear
   which are marked by black asterisks for the zone folded phonon lines and orange diamonds for the amplitude modes.  \textbf{c} and \textbf{d} Temperature dependent Raman spectra of \CVS in \Alg and \Ezg symmetry, respectively. For clarity, the spectra are consecutively offset by 0.5 cts/(s mW) each except for those measured at 9\,K. 
   }
   \label{fig:amplitude}
\end{figure*}

The weak additional lines observed below \Tcdw are indicated by black asterisks and orange diamonds in Fig.~\ref{fig:amplitude} \textbf{a} and \textbf{b}. In the zero-temperature limit the three \Alg lines are located at 43.0\,\wn, 105.4\,\wn and 200.0\,\wn. The six \Ezg lines appear at 43.2\,\wn, 60.0\,\wn, 101.0\,\wn, 180.0\,\wn, 208.2\,\wn and 224.0\,\wn. The lines at 43.0\,\wn and 105.4\,\wn are also found in pump-probe experiments \cite{Ratcliff:2021, Wangz:2021}. These new emerging lines are observed at nearly the same energies for different laser excitations (see Supplementary Materials D for details). Details of the temperature dependent positions and FWHMs of these lines can be found in Supplementary Materials E.

The lines marked by asterisks have weak and conventional temperature dependences and soften by less than 2\,\% between the low-temperature limit and \Tcdw (see Fig. \ref{fig:amplitude} \textbf{c} and \textbf{d}). The lines at 105\,\wn and 208\,\wn labeled with orange diamonds shift to lower energies by 17 and 10\,\wn, respectively, upon approaching \Tcdw, corresponding approximately to 15\,\% and 5\,\% relative shift. These two lines are identified as CDW AM in \CVS. The \Alg line broadens by approximately an order of magnitude close to \Tcdw and assumes a rather asymmetric shape in the range 40-80\,K (see Fig. \ref{fig:Tdependent} \textbf{a}).

\section{Discussion}


\subsection{Excitations across the gap}
The electronic Raman spectra (cf. Fig.~\ref{fig:cdwgap}) are consistent with a very large energy gap in excess of 80~meV. The ratio $2\Delta$/$k_{\rm B}$\Tcdw = 21-23, is almost 6 times the weak coupling mean field value of 3.53. This observation may be attributed to fluctuating short range CDW order \cite{Gruner:1994, Joshi:2019} which suppresses the observed transition below the mean-field value $T_{\rm MF}$ or to strong anharmonic electron-phonon coupling \cite{Varma:1983}. We carefully checked the electronic spectra above \Tcdw but could not find indications of fluctuations (see Supplementary Materials F for details) what favors strong coupling.

The anisotropy of the gap is substantial and manifests itself in the spectra as a broad intensity distribution and a small but significant difference in the intersection energies. It is not clear why the intensity amplitude (difference between spectra above and below \Tcdw) in \Alg symmetry is similarly strong as in \Ezg symmetry although only the \Ezg spectra project the $M$ points where a gap is expected. Given the complicated band structure of \CVS, it is possible that higher-order contributions to the \Alg vertex lend more weight to the $M$ regions than the lowest order vertex (compare Fig.~\ref{fig:lattice}~\textbf{d} and \textbf{e}). Alternatively, the band folding induced by the periodic distortion \cite{Kang:2022} could explain the relatively large \Alg amplitude (see the folded Brillouin zone in Fig.~\ref{fig:lattice}~\textbf{d}).



\subsection{New phonon lines below \Tcdw}

The additional phonon lines below \Tcdw result from the lowering of the lattice symmetry as spelled out by Wu \etal \cite{Wu:2022} and Liu \etal \cite{LiuG:2022} . The lattice distortion folds the phonon dispersion by a wave vector $\textbf{Q}$ that links the nesting points (M points in \CVS ) as seen in Fig.~\ref{fig:lattice}~\textbf{d}. Here, the phonons at the zone boundary are folded to the $\Gamma$ point and become Raman active \cite{Samnakay:2015, Albertini:2016}. If one considers a SoD or an iSoD distortion (see the insets of Fig.~\ref{fig:cdwgap} ~\textbf{c}) that are expected for \CVS (\cite{Christensen:2021, Liang:2021, Li:2021}), the V atoms move from 3$g$ (1/2, 0, 1/2) to 12$q$ ($x$, $y$, 1/2), and one expects eight additional Raman active modes (two in \Alg, four in \Ezg and two in \Elg). This figure matches the number of the new phonons in our measurements (asterisks only). Furthermore, most of the zone-folded modes quantitatively match the frequencies obtained in the DFT simulations when considering the iSoD distortion \cite{LiuG:2022}. Thus, the simulations of the phonon energies match the energy argument from the electronic spectra. The new lines have similarly large phonon-phonon coupling constants $\lambda_{\rm ph-ph}$ as the strong lines appearing above and below \Tcdw (see Fig. \ref{fig:renormalization} \textbf{d} and \textbf{f} and Tab. I in the Supplementary Materials E), indicating strong phonon-phonon coupling.

\subsection{Amplitude modes}
\begin{figure*}
   \includegraphics[width=\textwidth]{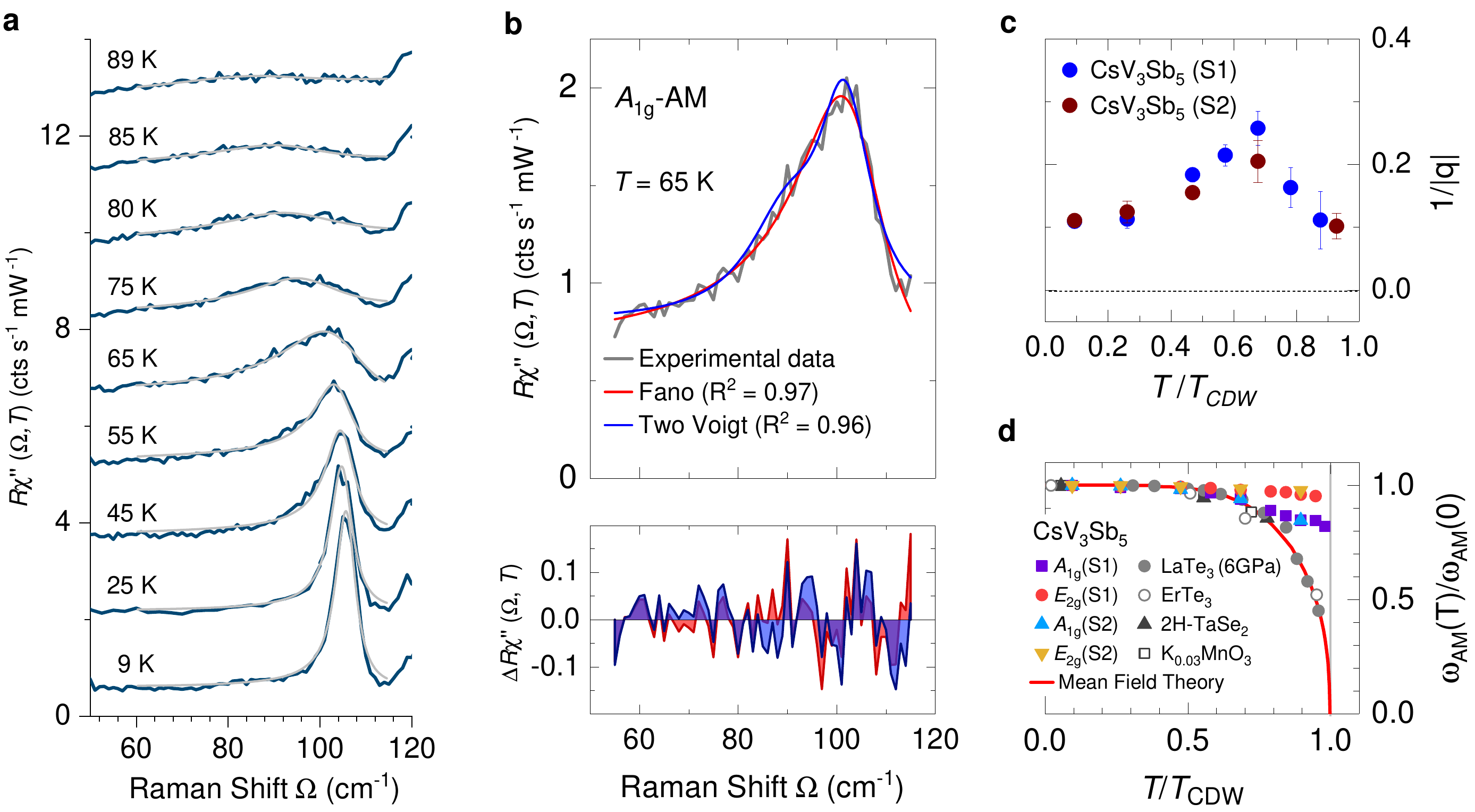}

   \caption{\textbf{Asymmetry of the $A_{1g}$ amplitude mode.} \textbf{a} The temperature evolution of the \Alg AM. Except for the spectrum at 9\,K, the data has been consecutively shifted by 1.5 cts/(s mW) for clarity. The Fano fits are superimposed in the spectra (light grey lines) \textbf{b} Top panel: The comparison between a superposition of two Lorentzian lines as suggested in Ref.~\onlinecite{Wu:2022} and a Fano line shape to the $A_{1g}$ AM. The Fano line yields a better residuum $R^2$. Bottom panel: The deviation of the spectra from the fitting curves (Red line: Fano, blue line: two Voigt).  \textbf{c} The asymmetry parameter $1/|q|$ as a function of normalized temperature in \CVS. \textbf{d} The temperature dependence of AM energies in various CDW materials. The data points in grey and black were extracted from Ref. \cite{Eiter:2013, Travaglini:1983, Hill:2019, Yumigeta:2022, Lavagnini:2010}. The peak energies of both AMs obivously deviate from the prediction of mean field theory (red curve) in \CVS. S1 and S2 in \textbf{c} and \textbf{d} denote two different samples.
   }
   \label{fig:Tdependent}
\end{figure*}

The lines at 105\,\wn and 208\,\wn in \Alg and \Ezg symmetry, respectively, have significantly stronger temperature dependences than the other lines appearing below \Tcdw, and are identified as AMs. Yet, the variation is much weaker than predicted by mean-field theory (see Fig. \ref{fig:Tdependent} \textbf{d}) and observed for the tritellurides, e.g. \cite{Lavagnini:2010,Eiter:2013}. There may be various reasons for the deviations: (i) Impurities lead to a saturation of the AM frequency at approximately the impurity scattering rate \cite{Devereaux:1993}. Here, this would imply a rather disordered system with an electronic mean free path of only a few lattice constants. (ii) An effect of strong electron-phonon coupling seems more likely, although enhanced coupling does not necessarily entail
a deviation from mean-field theory. Since Ginsburg-Landau theory \cite{Wu:2022} is applicable only close to the transition, where no data are available, the study of an extended temperature range below \Tcdw may be deceptive. In addition, the AM is not directly related to the gap, where single-particle (STS, ARPES) and two-particle (IR, Raman) results may return significantly different results. (iii) Strong phonon-phonon coupling and, consequently, higher order contributions from the phonons are not unlikely since the coupling $\lambda_{\rm ph-ph}$ of all modes below \Tcdw is substantial (see Fig.~\ref{fig:renormalization} \textbf{d} and \textbf{f} as well as Table I in Supplementary Materials E). This effect may enhance $2\Delta$/k$_{\rm B}$\Tcdw substantially and induce deviations from the mean-field temperature dependence of the AMs \cite{Varma:1983}. Thus, phonon-phonon coupling may contribute to the CDW formation here.

Particularly, the \Alg line at 105\,\wn exhibits anomalies in the line width and shape being incompatible with phonons. Right below \Tcdw the line width is as large as 50~\wn ( More details can be found in Supplementary Materials E). Previously the asymmetry has been interpreted in terms of two superimposed lines having individual temperature dependences \cite{Wu:2022}. We did not observe a double structure at low temperature but rather a narrow line having a width (FWHM) of approximately 6~\wn at 8\,K. We tested both hypotheses and find the Fano line to reproduce the data significantly better in the entire temperature range (see Fig. \ref{fig:Tdependent} \textbf{b}, more details can be found in Supplementary Materials H). Here, the \Alg AM line with energy $\omega_{\rm AM}$ and linewidth $\Gamma$ is fitted using the usual Fano function applicable for $\Gamma < \omega_{\rm AM}$,
\begin{equation}\label{eq:fano}
  I(\omega)=\frac{I_0}{|q^2+1|}\frac{(q+\varepsilon)^2}{1+\varepsilon^2};
  ~~~\varepsilon=2\frac{\omega-\omega_{\rm AM}}{\Gamma},
\end{equation}
to extract the asymmetry parameter $1/|q|$. We found that $1/|q|$ of the \Alg AM becomes maximal at 68~K where the putative transition from $2\times2\times2$ to $2\times2\times4$ stacking is expected \cite{Stahl:2022} (see Fig. \ref{fig:Tdependent} \textbf{c}).

The Fano line of the AM is unique in \CVS  and has not been observed in other well-known CDW materials, such as ErTe$_3$ \cite{Eiter:2013}, LaTe$_3$ \cite{Lavagnini:2010}, 2H-TaSe$_2$ \cite{Hill:2019}, and K$_{0.3}$MoO$_3$ \cite{Travaglini:1983}, where all AMs have a symmetric Lorentzian line-shape (See Supplementary Materials I). The asymmetric AM, along with the missing soft mode behaviour in the acoustic branches \cite{Li:2021,Xie:2022} and the large $2\Delta$/k$_{\rm B}$\Tcdw ratio argue against the weak coupling picture. Thus, strong coupling involving phonon-phonon and electron-phonon interaction is presumably the driving force behind the formation of the CDW in \CVS. In this case, the conduction electrons strongly couple with the involved phonons, and may further interact with the AM inducing the Fano line shape. The decrease of the $1/|q|$ towards zero temperature is presumably a result of the CDW gap opening below 1500~\wn, evidenced by the reduction of scattering intensity below some 1500~\wn. In addition, we derive signatures of strong phonon-phonon coupling from the anharmonic decay of the Raman-active optical modes giving another boost to the CDW formation \cite{Varma:1983}.\\

\section{Conclusions}
We performed a polarization- and temperature-dependent Raman scattering study of the Kagome metal \CVS. The electronic continua in both the \Alg and \Ezg symmetry exhibit a spectral-weight redistribution below the charge-density-wave transition temperature, \Tcdw $\sim 95$~K, in agreement with DFT simulations. This redistribution indicates an energy gap of $2\Delta\lesssim 1500$~\wn (185~meV) corresponding to $2\Delta/k_BT_{\rm CDW}$ of approximately 22. The DFT results favour an iSoD distortion for $T\to 0$. In the low-energy part of the spectra several phonons pop out below \Tcdw in addition to the two modes in \Alg and \Ezg symmetry Raman active at all temperatures. The additional lines are related to the lattice distortion due to the CDW transition. Intriguingly, we identified two CDW amplitude modes having energies of $\omega^{A{1g}}_{AM}=105$~\wn and $\omega_{AM}^{E{2g}}=208$~\wn in the low-temperature limit. The \Alg AM  couples strongly to a continuum as indicated by the Fano-type line shape displaying the strongest asymmetry at the putative cross-over temperature of approximately 60~K between $2\times2\times2$ to $2\times2\times4$ ordering \cite{Stahl:2022}. The mode's temperature dependence is weaker than predicted by mean field theory. This discrepancy may result from either impurities \cite{Devereaux:1993} or strong coupling \cite{Varma:1983}. Since the crystals are well-ordered we consider the strong-coupling scenario including anharmonic phonon-phonon and electron-phonon coupling \cite{Varma:1983} more likely. This interpretation is consistent with the large electronic gap and the asymmetric AM. Thus, the co-operation of strong electron-phonon and phonon-phonon coupling may be more likely a route to the CDW transition in \CVS than, e.g., nesting.\\

\noindent\textbf{Methods}\\
\noindent\textbf{Samples:} Single crystals of \CVS were grown from liquid Cs (purity 99.98\,\%), V powder (purity 99.9\,\%) and Sb shot (purity 99.999\,\%) via a modified self-flux method \cite{Ni:2021}. The mixture was put into an alumina crucible and sealed in a quartz ampoule under argon atmosphere. The mixture was heated at 600\,C for 24\,h and soaked at 1000\,C for 24\,h, and subsequently cooled at 2\,C/h. Finally, the single crystals were separated from the flux by an exfoliation method. Apart from sealing and heat treatment procedures, all other preparation procedures were carried out in an argon-filled glove box. The crystals have a hexagonal morphology with a typical size of $2\times2\times1$\,mm$^3$ and are stable in the air. The sample used for the Raman experiments has a \Tcdw of 95\,K, characterized by resistivity and in-plane magnetic susceptibility (see Supplementary Materials A for details).\\

\noindent\textbf{Light scattering:} The inelastic light scattering experiments were preformed in pseudo-backscattering geometry. The samples were mounted on the cold finger of a $^4$He flow cryostat immediately after cleaving. For excitation, a solid-state and an Ar$^{+}$ laser emitting at 575\,nm, 514\,nm, and 476\,nm were used. The inelastic spectra were divided by the Bose factor yielding $R\chi^{\prime\prime}(\Omega,T) = \pi\{1+n(\Omega,T)\}^{-1}S(q \approx 0, \Omega)$ where $\chi^{\prime\prime}$ is the imaginary part of Raman response function, $R$ is an experimental constant, and $S(q \approx 0, \Omega)$ is the dynamical structure factor \cite{Devereaux:2007}.  Typical phonon lines are describe by Lorentzians. If the width is close to the spectral resolution or below a Voigt function (convolution of a Lorentzian and a Gaussian) has to be used.

\begin{figure*}[ht]
   \includegraphics[width=\textwidth]{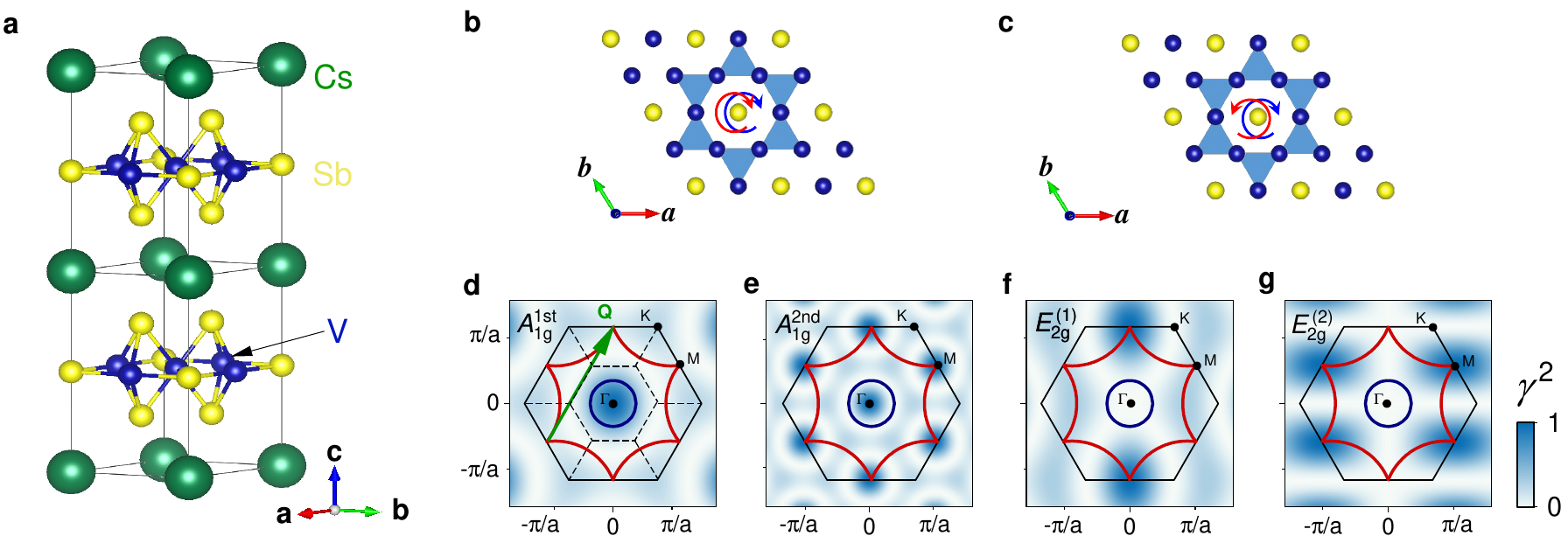}
   \caption{\textbf{Structure and polarization configurations in \CVS with the related Raman vertices.} \textbf{a} The crystal structure. Cs, V, and Sb atoms are shown in green, blue and yellow, respectively. \textbf{b} and \textbf{c} Kagome lattice of the V-Sb layers. The polarization configurations of \Alg and \Ezg symmetries are superimposed as blue and red circular arrows. The Raman vertices are shown with the color mapping for the \textbf{d} first- and \textbf{e} second-order \Alg symmetry, \textbf{f} and \textbf{g} first order \Ezg symmetry. The first Brillouin zone is represented by the black hexagon.  The red (hole pockets) and blue (electron pocket) curves indicate the Fermi pockets. The green arrow in \textbf{d} illustrates the ordering vector \textbf{Q} that connects two $M$ points. The dashed lines indicate the corresponding folded Brillouin zone.   }
   \label{fig:lattice}
\end{figure*}

For projecting the \Alg and \Ezg symmetries $RR$ and $RL$ polarization configurations were used, respectively. In terms of perpendicular linear polarizations $x$ and $y$, $R$ and $L$ are given by $R=\frac{1}{\sqrt{2}}(x+iy)$ and $L=\frac{1}{\sqrt{2}}(x-iy)$, respectively. The configurations with respect to the Kagome plane are shown in Fig.~\ref{fig:lattice}~\textbf{b} and \textbf{c}. For electronic Raman scattering the form factors are important and highlight parts of the Brillouin zone. The form factors or Raman vertices may be expressed in terms of the band curvature or crystal harmonics \cite{Devereaux:2007}. The first- and second-order crystal harmonics of \Alg symmetry and the first order crystal harmonics of \Ezg symmetry and the position of the Fermi pockets of \CVS are shown in Fig.~\ref{fig:lattice}~\textbf{d}-\textbf{f} and illustrate the sensitivity of the experiment (The vertices derived from the crystal harmonics functions can be found in Supplementary Materials B. For details see Ref. \cite{Mijin:2021}). \\

\noindent\textbf{DFT simulations:} DFT calculations were performed using VASP \cite{Kresse:1993} with plane wave augmented (PAW) pseudopotentials and a 300 eV energy cutoff. In all calculations the pristine, SoD, and iSoD states were considered independently as 2$\times$2 distortions. Lattice constants were calculated with pristine structures and kept fixed in CDW states. Minimum energy CDW states were found around 1.5$\%$ lattice distortion from the pristine structure. Structural and electronic calculations were performed on each of the three states. A 17$\times$17$\times$9 KPOINT grid was used for electronic calculations. The electronic response in the main text was approximated using the joint density of states with the following equation:

\begin{equation}
    \begin{split}
\chi^{''}_{\mu\nu}(\Omega)=\sum_k\gamma_k^\mu\gamma_k^\nu~~~~~~~~~~~~~~~~~~~~~~~~~~~~~~~~~~~~~\\
\times\int d\omega A^\mu(k,\omega)A^\nu(k,\omega+\Omega)[f(\omega)-f(\omega+\Omega)]
    \end{split}
\end{equation}

where
\begin{equation}
   A^\mu(k,\omega)=\frac{1}{\pi}\frac{\Gamma}{\Gamma^2+(\omega-\epsilon^\mu_k)^2}
\end{equation}
In this equation, $\mu$ and $\nu$ are band indexes, $\gamma_k^\mu$ is the Raman vertex at momentum $k$ and band $\mu$, $A(k,\omega)$ is the spectral weight at momentum $k$ and energy $\omega$, $\Omega$ is the Raman shift energy, $f(\omega)$ is the Fermi-Dirac function, $\epsilon_k^\mu$ is the band dispersion, $\Gamma$ is the energy broadening. A broadening of 0.02 eV was used in the results in the main text. In the calculations, the Raman vertex $\gamma_k^\mu$ is fixed at a value of 1. Thus, selection rules are ignored for the time being.

Phonon calculations were performed using the Phonopy code package \cite{Togo:2023, phonopy}. A 3$\times$3$\times$4 K-mesh was utilized for the calculations. Our primary focus here is on the energy of the \Alg phonon. In the pristine state, we observed a good agreement between the calculated phonons and those observed experimentally. Based on success by previous studies \cite{Tan:2021}, for calculations comparing the pristine and CDW states, we employed DFT-3 to stabilize the phonon frequencies in the distorted phases, resulting in an overall frequency shift towards higher energy while maintaining reliable relative positions. Electron phonon coupling to the $A_{1g}$ and $E_{2g}$ was calculated using the frozen phonon method and found to be negligible.\\

\noindent\textbf{Acknowledgments}\\
We thank R.-Z. Huang for fruitful discussions. This work is supported by the Deutsche Forschungsgemeinschaft (DFG) through the coordinated programme TRR80 (Projekt-ID 107745057) and projects HA2071/12-1 and -3. L.P. and R.H. were partially supported by the Bavaria-California Technology Center (BaCaTeC) under grant number A3 [2022-2]. G.H. would like to thank the Alexander von Humboldt Foundation for a research fellowship. The work in China was supported by grants from the National Key Research and Development Projects of China (2022YFA1204104), the Chinese Academy of Sciences (ZDBS-SSW-WHC001 and XDB33030000),and the National Natural Science Foundation of China (61888102). Theory work at SLAC National Accelerator (EFC, TPD) was supported by the U.S. Department of Energy, Office of Basic Energy Sciences, Division of Materials Sciences and Engineering, under Contract No. DE-AC02-76SF00515. The computational results utilized the resources of the National Energy Research Scientific Computing Center (NERSC) supported by the U.S. Department of Energy, Office of Science, under Contract No. DE-AC02-05CH11231. \\

\noindent\textbf{Author contributions}\\
R.H. and G.H. conceived the project.
G.H., L. P., D. L. and R.S. performed the Raman measurements. G.H., L.P., E.C., T.P.D., and R.H. analysed the Raman data. Z.Z., H.-T.Y. and H.-J.G.
synthesised and characterised the samples. E.C., B.M. and T.P.D. performed
DFT calculations. G.H., L.P., E.C. and R.H. wrote the manuscript with comments
from all the authors.\\

\noindent\textbf{Competing interests}\\
The authors declare no competing interests.
\bibliography{refs}

\newpage

\end{document}